\documentclass[fleqn,a4paper]{elsart}

\newcommand{\im}{\mathop\mathrm{Im}}

\usepackage{amsmath}
\usepackage{amssymb}

\begin{document}

\begin{frontmatter}
\title{On  the uncertainty relations for angular observables and beyond them}
\author{S.Dumitru}\footnote{Electronic mail: s.dumitru@unitbv.ro}
\address{Department of Physics, ``Transilvania'' University, B-dul Eroilor 29, 
R-2200 Bra\c{s}ov, Romania}
\date{\today}
\maketitle
\begin{abstract}
For angular observables pairs (angular momentum-angle  and number-phase) the adequate 
reference element of normality is not the Robertson-Schr\"{o}dinger     uncertainty relation
but a Schwarz formula regarding the quantum fluctuations. Beyond such a fact the 
traditional interpretation of the uncertainty relations appears as an unjustified doctrine.
\end{abstract}
\end{frontmatter}

\section{Introduction}

The problem of uncertainty relations (UR) for angular variables regards the
anomalies of the pairs $L_{z}-\varphi$ 
( angular momentum - azimuthal angle)
and $N-\phi$ (number - phase) in respect with the predominant conceptions
about quantum mechanics (QM). It was brought forward as a burning question
four decades ago \cite{1,2,3} and since then it is known as subject of 
many debates
(see the review works \cite{4,5,6,7,8,9} and references. 
But in spite of such a history
the mentioned problem still remains \cite{9,10,11} as an open question 
without an agreement of opinions, because the existing approaches are dissimilar both
quantitatively and qualitatively. Then the searches for new and deeper
approaches of the same problem and related facts are topical. One of the
searched approaches can be obtained by investigating the identity and
viability of normality reference element comparatively with which the
mentioned anomalies are evaluated. Such an investigation together
with some connected questions define the goal of the present paper. We shall
show that the Robertson Schr\"{o}dinger UR (RSUR), currently assumed as the
normality reference element, is inadequate in respect with the angular
observables. Also we find that the respective observables can be described
without problems by the usual procedures of QM. Then a simple relation of
Schwartz type appears in the posture of a true reference element of normality. The
mentioned rebutment  of RSUR suggests directly a reconsideration of the
traditional interpretation of UR (TIUR). Here we shall present such a
reconsideration by developing some incipient ideas from our works \cite{12,13,14,15,16}.
Therethrough we find that TIUR is an unjustified doctrine while UR require a
natural reinterpretation. So UR are found to be not crucial physical
formulas but simple fluctuations relations with natural analogues in
non-quantum physics. Such findings give a justification, from another
perspective, of the observation that \cite{17} : \emph{"there is nothing especially 
quantum
mechanical about the ...\emph{'coordinate momentum UR'}.. .  \textbf{per se}"}. 
Also the new
regard about  UR induce the idea that the natural description of
quantum measurements has to be separated from the objectives of usual QM

\section{The RSUR as an incorrect reference element}

Currently (see \cite{1,2,3,4,5,6,7,8,9} and quoted publications) one assumes that for 
$L_{z}-\varphi$ and $N -\phi$ pairs the alluded
reference element of normality is given by the RSUR
\begin{equation}\label{eq:1}
\Delta A \cdot \Delta B \ge \frac{1}{2}\,\vert \langle 
[ \hat A ,\hat B ] \rangle \vert 
\end{equation}
Here $\Delta$A and $\Delta$B denote the standard deviations of
observables A and B while $ [ {\hat A,\hat B} ] $ signifies the commutator 
of $\hat A $ and $\hat B$ respectively
$\left\langle {...} \right\rangle $ represent the mean value. According
to the usual procedures of QM the observables 
$ L_{z} - \varphi  $ should be
described by the conjugated operators
\begin{equation}\label{eq:2}
\hat L_{z}  =  - i\hbar \;\frac{\partial }{{\partial \varphi }}
\quad \quad and \quad \quad \hat\varphi = \varphi \, \cdot 
\end{equation}
respectively by the commutation relation
\begin{equation}\label{eq:3}
[\hat L_{z}, \hat \varphi] = -i\hbar
\end{equation}
 So for   $L_{z}-\varphi $ pair the 
 RSUR (2.1) requires directly the relation
 \begin{equation}\label{eq:4}
\Delta L_z  \cdot \Delta \varphi \ge \frac{\hbar}{2}
\end{equation}
The anomaly of $L_z-\varphi$ pair appears as the incorrectness of the
relation \eqref{eq:1} in respect with the usually quoted physical situations. In
fact the alluded situations regard exclusively the restricted class of
circular  rotations (CR). Such CR are specific for a particle on a circle, a
2-D rigid  rotator and non-degenerate spatial rotations. One finds examples of systems
with spatial  rotations  in cases of a particle on a sphere, of a
3-D rigid rotator  and an electron in a hydrogen like atom. The respective
rotations  are considered as non-degenerate if all the implied quantum 
numbers have unique values.

In all the cases of CR the part of the wave function important 
for $L_{z}-\varphi$ pair has the form
\begin{equation}\label{eq:5}
\psi_m (\varphi) = (2\pi )^{- 1/2} e^{i m \varphi }
\end{equation}
with 
\begin{equation*}
\varphi \in [0,2\pi),\qquad  \psi (2\pi) :=  
\lim\limits_{\substack{\varphi\to 2\pi \\ \varphi < 2\pi}} \psi (\varphi)
\end{equation*}
 and a single 
value for the integer number $m$. Then in the respective cases one obtains 
the expressions

\begin{equation}\label{eq:6}
\Delta L_z  = 0 \, ,  \qquad 
\Delta \varphi  = \frac{\pi }{{\sqrt 3 }}
\end{equation}
But these expressions are incompatible with the relation \eqref{eq:4}

In order to avoid the mentioned incompatibility many publications 
promoted the conception that for  the 
$ L_{z}  -  \varphi  $ pair the usual
procedures of QM do not work correctly. Consequently  the idea that the formula 
\eqref{eq:4} must be prohibited and replaced by adjusted
$\Delta L_z  - \Delta \varphi$ relations resembling with RSUR \eqref{eq:1} was 
accredited.  So, along the years, a lot of such adjusted relations were promoted. 
In the main the respective relations are expressible in one of the following forms
\begin{equation}\label{eq:7}
\Delta L_z  \cdot \Delta f(\varphi )\ge \hbar  \left\langle g(\varphi ) \right\rangle 
\end{equation}
\begin{equation}\label{eq:8}
(\Delta L_z  )^2  + \hbar^2  (\Delta u(\varphi ))^2  \ge \hbar^2 \left\langle 
 v(\varphi ) \right\rangle ^2 
\end{equation}
\begin{equation}\label{eq:9}
\Delta L_z  \cdot \Delta  \varphi  \ge  \frac{\hbar }{2}\left\vert 1  - 
2\pi \left\vert  \psi (2\pi ) \right\vert^2  \right\vert
\end{equation}
In \eqref{eq:7}-\eqref{eq:9} $f(\varphi)$, $g(\varphi)$, $u(\varphi)$and $v(\varphi)$ 
denote various adjusting functions of $\varphi$ introduced by means of some 
circumstantial considerations.

A minute examination of the facts shows that, in essence, the set of the relations 
\eqref{eq:7}-\eqref{eq:9} is affected by the following shortcomings (\textbf{Shc}):
\begin{itemize}
\item \textbf{Shc.1}: None of the respective relations  is agreed unanimously as a correct
 $\mathop {\Delta L}\nolimits_z  - \mathop {\Delta \varphi }\nolimits_{} $
relation able to replace the formula \eqref{eq:4}
\item \textbf{Shc.2}: Mathematically the alluded relations are not mutually
equivalent.
\item \textbf{Shc.3}: The relations \eqref{eq:7}-\eqref{eq:8} do not have rational supports 
in the usual formalism of QM (that however works very well in a huge number 
of applications. 
\item \textbf{Shc.4}: The considerations appealed in the promotion of relations
 \eqref{eq:7}-\eqref{eq:8} are not based on natural physical arguments 
\end{itemize}

\emph{\underline{Observation} }: We do not associate the formula \eqref{eq:9} with \textbf{Shc.3-4}
because it is justifiable within the usual framework of QM (see below the relations
\eqref{eq:22} and \eqref{eq:31}.

In the context of the above discussions  another fact connected 
with the relation \eqref{eq:4} is of interest. In spite of the known idea that it must be 
prohibited the respective relation appears as valid in some nontrivial physical 
situations regarding the non-circular rotations (NCR) By NRC we refer to the 
quantum torsion pendulum (QTP) respectively to the degenerate spatial rotations
(of the above mentioned systems). A rotation (motion)is degenerate if the energy 
of the system is well precised while the non-energetic quantum numbers take all the 
permitted values.

From the class of NCR  let us firstly refer to the case of a QTP which \cite{14,15}  is
nothing but a harmonic oscillator characterized by the Hamiltonian 
\begin{equation*}
\hat H\; = \; - \,\frac{{\hbar ^2 }}{{2I}}\,\frac{{\partial ^2 }}
{{\partial \varphi ^2 }}\; + \;\frac{{I\omega ^2 }}{2}\,\varphi ^2 
\end{equation*}
 with $\varphi \in (-\infty, +\infty)$,
$I$ = moment of inertia and $\omega$ = angular frequency. In the state with the energy
$ E_{n} = \hbar \omega \left( {n + \frac{1}{2}} \right)$ 
the QTP is described by the wave function
\begin{equation}\label{eq:10}
\psi _n (\varphi ) = \psi _n (\xi ) \propto \exp \left\{  - \frac{\xi ^2 }{2}\right\}
 H_n (\xi ),\quad \xi  = \varphi \sqrt {\frac{{I\omega }}{\hbar }} 
\end{equation}
Here : $n = 0,1,2,... $= vibrational quantum number, $H_n(\xi )$= Hermite polinomials of
$\xi$. For observables $L_z$ and $\varphi$ described by operators \eqref{eq:2} by means of   
\eqref{eq:10} one obtains
\begin{equation}\label{eq:11}
\Delta L_{z}  = \sqrt {\hbar I\omega \left(n + \frac{1}{2}\right)} \, ,\quad \Delta \varphi 
 = \sqrt {\frac{\hbar }{{I\omega }}\left(n + \frac{1}{2}\right)} 
\end{equation}
With these expressions one finds that for QTP the pair $L_z -\varphi$
satisfy the prohibited formula \eqref{eq:4}.

From the same class of NRC now let us refer to the cases of degenerate spatial rotations
regarding a particle on a sphere or a 3D rigid rotator. In such cases the energy
$E=\hbar ^2 l(l + 1)/2I$ and orbital number $l$ have well-precised values while 
the magnetic number $m$ takes the values $ -l, -l+1,...,-1, 0, 1, ...,l-1, l$. 
The corresponding wave functions for the considered cases has the form :
 \begin{equation}\label{eq:12}
\psi _l (\varphi )=\sum\limits_{m =  - l}^l {c_m } Y_{lm} (\vartheta ,\varphi )
\end{equation}
where $Y_{lm} (\vartheta ,\varphi )$ = spherical functions and $c_m$ = complex 
coefficients which satisfy the condition $\sum_{m=-l}^l {\left| {c_m } \right|} ^2 =1$.\\
With $L_z$ and $\varphi$ described by the operators \eqref{eq:2}and using  \eqref{eq:12}
one obtains  \begin{equation}\label{eq:13}
\left( \Delta L_z  \right)^2 =\sum_m =  - l^ l \left| c_m  
\right|^2  \hbar^2 m^2  - \left[ \sum_{m=-l}^l c_m \hbar m  \right]^2
\end{equation}

\begin{eqnarray}\label{eq:14}
(\Delta\varphi)^2 &=& \sum_{m=-l}^{l} \sum_{m'=-l}^{l} c_m^* c_{m'} 
(Y_{lm},\varphi^2 Y_{lm'}) - \nonumber \\
 &&- \left [ \sum_{m=-l}^{l}\sum_{m'=-l}^{l} c_m^* c_{m'}  
 (Y_{lm},\varphi Y_{lm'})\right]^2
\end{eqnarray}
where $(f,g)$ denotes the scalar product of the functions $f$ and $g$.

With \eqref{eq:13} and \eqref{eq:14} one finds that in the cases described by 
\eqref{eq:12} it is possible that the prohibited formula \eqref{eq:4} to be verified. 
The respective possibility is conditioned by the concrete values of the coefficients $c_m$.

Now let us refer to the pair $N-\phi$ (number-phase) which \cite{4,5,6,7,8,9} was also found  
as showing an anomaly in respect with the same reference element given by RSUR \eqref{eq:1}.
The mentioned pair refers to a quantum oscillator described by a wave function $\psi_n$
like \eqref{eq:10} (with $\xi =x\sqrt {\frac{{M\omega }}{\hbar }} $ in the case of 
a recti-liniar oscillator of mass $M$ and Cartesian coordinate $x$ ). The corresponding 
operators $\hat N$ and $\hat \varphi$ are introduced (according to a Dirac's idea) by the relations
\begin{equation}\label{eq:15}
\hat a=e^{i\hat \phi } \sqrt {\hat N} \, ,\qquad 
\hat a^ +  =\sqrt {\hat N} e^{ - i\hat \phi } 
\end{equation}
where ${\hat a}$ and $\hat a^+$ are the known ladder (annihilation and creation) operators. 
From \eqref{eq:15} it follows directly the commutation formula
\begin{equation}\label{eq:16}
[ {\hat N\;,\;\hat \varphi } ]=i
\end{equation}
Then RSUR \eqref{eq:1} should imply the relation
\begin{equation}\label{eq:17}
\Delta N \cdot \Delta \phi \ge \frac{1}{2}
\end{equation}
On the other hand, because  in the considered case $\phi \in [0,2\pi)$
and $\psi_n$ is an eigenfunction of ${\hat N}$, one obtains
\begin{equation}\label{eq:18}
\Delta N=0 \, ,\qquad \Delta \phi \le 2\pi 
\end{equation}
But such values for$\Delta N$ and $\Delta \phi$ are in evident discordance with 
\eqref {eq:17} and so they reveal the anomaly of the $N-\phi$ pair in respect with 
RSUR \eqref {eq:1} as reference element. We add here the observation that, in fact , 
the pair $N - \phi$ is in a situation completely similar with that of the pair 
$L_z - \varphi$ in CR cases. The respective similarity can be pointed out as follows.
 If the wave functions are taken in the $\phi$ - representation from \eqref{eq:16}
 it results that the operators ${\hat N}$ and ${\hat \phi}$ have the expressions
 \begin{equation}\label{eq:19}
 \hat N=i\,\frac{\partial }{{\partial \phi }}\quad ,\quad \quad 
 \quad \hat \phi =\phi \cdot 
\end {equation}
Then the Schr\"{o}dinger equation for the oscillator take the form
\begin{equation}\label{eq:20}
\hbar \omega (i\frac{d}{{d\phi }}+ \frac{1}{2})\psi =E \psi 
\end {equation}
By considering $\psi_n (2\pi)=\psi_n(0)$ and $E \geq 0$ from \eqref{eq:20}one obtains
\begin{equation}\label{eq:21}
\psi_{\rm n} =\frac{1}{{\sqrt {2\pi } }}\,e^{ - in\phi } 
\end{equation}
respectively $E_{n\;} =\hbar \omega \,(n\; + \;\frac{1}{2})$ and $n = 0,1,2,...  $ .
So the couples of relations \eqref{eq:3}/\eqref{eq:16} and \eqref{eq:5}/\eqref{eq:21}
attest the announced mathematical similarity between the pairs $N-\phi$ and
$L_z-\varphi$.

The alluded circular similarity is evidenced also by the various adjusted 
$\Delta N-\Delta \phi$ formulas proposed in the literature 
(see the works \cite{4,5,6,7,8,9} and references) in order to replace \eqref{eq:17} 
and to avoid the anomaly of $N - \phi$ pair in respect with RSUR \eqref{eq:1}. 
In their essence the respective formulas are completely analogous with the relations  
\eqref{eq:7}-\eqref{eq:9} for $L_z - \varphi$pair. Moreover , it is easy to see, that the alluded 
$\Delta N - \Delta \phi$ formulas 
are affected by shortcomings which are similar with the above mentioned \textbf{Shc.1-4}.
Here it is important to remark that the above mentioned shortcomings, for both pairs 
$L_z - \varphi$ and $N - \phi$ , have an unavoidable character. This means that RSUR 
\eqref{eq:1}taken as normality reference element implies inevitable anomalies for  the 
respective pairs in the CR cases. On the other hand, as it was shown above, RSUR 
\eqref{eq:1} can not offer a basis for an unitar approach of all cases (of CR and NCR type)
regarding the $L_z - \varphi$ pair. Then one can conclude that , in fact for angular
observables  $L_z - \varphi$ and $N - \phi$, the RSUR \eqref{eq:1}is not adequate for
  the role of reference element. In the next section we shall show that in usual QM one
 can find an adequate candidate for such an role.
 \section {The true reference element} 
 Now we shall investigate the mentioned inadequacy of RSUR \eqref {eq:1} by searching its 
 true origin and validity conditions. Such a search can be done as follows with the aid of 
 some elements/notations from usual QM.
 
 We consider a quantum system in a state described by the wave function $\psi$. The
  observables $A_j (j=1,2,..,r)$of the system are associated with the operators $\hat A_j$.
If $(f,g)$ denote the scalar product of two functions $f$ and $g$, for two observables 
$A_1 = A$ and $A_2 = B$ one can write the following Schwarz relation
 \begin{equation}\label{eq:22}
(\delta \hat A\psi ,\delta \hat A\psi ) \cdot (\delta \hat B\psi ,\delta \hat B\psi )
\; \ge \;| \;{(\delta \hat A\psi ,\delta \hat B\psi )}\; |^2 
\end {equation}
where $\delta \hat A=\hat A - \left\langle A \right\rangle$ and
$\left\langle A \right\rangle  = (\psi ,\hat A\psi )$= the mean (expected)value of
the observable A. According to the usual rules of QM 
$(\delta \hat A\psi ,\delta \hat A\psi )^{\frac{1}{2}} =\Delta A\; = $ standard
deviation of A. So from \eqref{eq:22} one obtains directly
\begin{equation}\label{eq:23}
\Delta A \cdot \Delta B\; \ge \;| \;{(\delta \hat A\psi ,\delta \hat B\psi )}\; |
\end {equation}
This Schwarz formula is generally valid for any wave function $\psi$ and any observables
$A$ and $B$. It implies the less general relation which is RSUR \eqref{eq:1} only when 
the operators $\hat A = \hat A_1$ and $\hat B = \hat A_2$ satisfy the conditions :
\begin{equation}\label{eq:24}
(\hat A_j \psi ,\hat A_k \psi )=(\psi ,\hat A_j \hat A_k \psi )\quad 
\quad (j = 1,2;\;k = 1,2)
\end {equation}
Indeed when \eqref {eq:24} are satisfied one can write
\begin{equation}\label{eq:25}
(\delta \hat A\psi ,\delta \hat B\psi )=\frac{1}{2}(\psi ,
(\delta \hat A\delta \hat B + \delta \hat B\delta \hat A)\psi ) - 
\frac{i}{2}(\psi ,i\,[ {\hat A,\hat B} ]\,\psi )
\end {equation}
where the two terms from the right side are purely real respectively purely imaginary 
quantities. When \eqref{eq:24} are satisfied the formula \eqref{eq:23} gives directly 
the RSUR
\begin{equation}\label{eq:26}
\Delta A\, \cdot \Delta B\; \ge \;\frac{1}{2}\;| \;{\langle \,{[ 
{\hat A,\hat B} ]} \,\rangle }\; |
\end{equation}
The above presented considerations justify the following observations :
\begin{itemize}
\item \textbf{(i) }The Schwarz formula \eqref{eq:23} is aboriginal in respect with  the RSUR
\eqref{eq:26}. Moreover it is \eqref{eq:23} always valid, independently if the conditions 
\eqref{eq:24} are satisfied or no.
\item \textbf{(ii)} The RSUR \eqref{eq:26}/\eqref{eq:1} is valid only in the circumstances 
strictly delimited by the conditions \eqref{eq:24} and it is false in all other
 situations.
 \end{itemize}

The noted observations suggest to investigate if the previously discussed behavior 
of RSUR in respect with the the pairs $L_z - \varphi$ and $N - \phi$ can be correlated 
with the conditions \eqref{eq:24}. For such an investigation the following facts are of 
direct interest. In the cases described by the wave functions \eqref{eq:5} and 
\eqref{eq:21} for  $L_z - \varphi$ and $N - \phi$ one finds respectively
\begin{equation}\label{eq:27}
(\hat L_z \psi _m ,\;\hat \varphi \psi _m )\;=\;(\psi _m ,\;\hat L_z 
\hat \varphi \psi _m )\; - \;i\hbar
\end{equation}
\begin{equation}\label{eq:28}
(\hat N\psi _n ,\hat \phi \psi _n )=(\psi _n , \hat N \hat \phi \psi _n ) + i
\end{equation}
For the pair $L_z-\varphi$ in the cases associated with the wave functions 
\eqref{eq:10} respectively \eqref{eq:12} one obtains:
\begin{equation}\label{eq:29}
(\hat L_z \psi _n ,\hat \varphi \psi _n )=(\psi _n ,\hat L_z \hat \varphi \psi _n )
\end{equation}
\begin{eqnarray}\label{eq:30}
 (\hat L_z \psi _l ,\hat \varphi \psi _l ) &=&(\psi _l \;,\;\hat L_z 
 \hat \varphi \psi _l )+ \nonumber \\ 
 && + i \hbar \left\{ 1 + 2 \,\im \left[\sum_{m=-l}^l 
 \sum_{m'=-l}^l c_m^* c_{m'} m  (Y_{lm} ,\hat \varphi Y_{lm'} )
 \right] \right\} 
\end{eqnarray}
Relations \eqref{eq:27}-\eqref{eq:30} justify the following remarks. RSUR \eqref{eq:26}/
\eqref{eq:1} is essentially inapplicable for the pairs $L_z-\varphi$ and $N-\phi$
in the cases of CR described by \eqref{eq:5} and \eqref{eq:21}. In respect with 
\eqref{eq:10}the RSUR \eqref{eq:26}/\eqref{eq:1}is always applicable. In the situations
associated with \eqref{eq:12} the applicability of RSUR \eqref{eq:26}/\eqref{eq:1}
to the $L_z-\varphi$ pair depends on the values of the second term from right side of 
\eqref{eq:30}. It is important that in all cases regarding the pairs $L_z-\varphi$ 
and $N-\phi$ the Schwarz formula \eqref{eq:23} remains valid. In the above noted 
situations when \eqref{eq:24} are not satisfied the respective formula degenerate 
into the trivial equality $0=0$.

Here is the place to  mention also the fact that, for any wave function $\psi(\varphi)$
with $\varphi\in[0, 2\pi)$ and $\psi(2\pi) = \psi(0)$,  the relation :
\begin{equation}\label{eq:31}
| (\delta \hat L_z \psi ,\delta \hat \varphi \psi ) | 
\ge \frac{\hbar }{2}\left| 1- 2\pi \left| \psi (2\pi ) \right| \right|
\end{equation}
is generally true. 
This result shows that the adjusted relation \eqref{eq:9} is only a secondary piece 
derivable from the general Schwarz formula \eqref{eq:23}.

The facts pointed out in this section prove that , in respect with the 
pairs $L_z-\varphi$ 
and $N - \phi$ the RSUR \eqref{eq:1} is not adequate for the role of normality 
reference element.
It also  results that for such a role  the Schwarz formula
\eqref{eq:23} is adequate without problems. The respective formula is valid for 
 any state (wave function) and for 
all pairs of observables (particularly for $L_z - \varphi$ and $N - \phi$). So one finds 
that in reality the usual procedures of QM work well and without anomalies in all 
situations of interest for physics.

Previous findings show that for $L_z$ and $\varphi$ the relations \eqref{eq:2} -
 \eqref{eq:3}are always viable. So in a natural framework of QM it is not necessary to 
 replace  the respective relations with some substitutions (like \cite{18}: 
  $\hat L_z =
  - i\hbar  \frac{\partial }{{\partial \varphi }} + \alpha $, or \cite{6}:  
  $[\hat L_z , \hat \varphi  ]= - i\hbar  + \delta\, $ ($\delta$ = Dirac's function)).
  
\emph{\underline{Observation}:}	The deadlock of RSUR in respect with the pairs $L_z - \varphi$ and
$N - \phi$ is directly connected with the conditions \eqref{eq:24}. Then it is strange 
that in almost all the QM literature the respective conditions are ignored. The reason 
seems to be related with the fact that in Dirac's $\left\langle {bra} \right|$  and
 $\left| {ket} \right\rangle$ notations  (which dominate in the nowadays publications)
 the terms from the both  sides of \eqref{eq:24}have a unique notation - namely 
 $\langle {\psi | {\hat A_j \hat A_k } |\psi } \rangle $. Such a 
 uniqueness in notations can induce the confusion (unjustified supposition) that the  
 conditions \eqref{eq:24}are always satisfied. It is interesting to note that systematic 
 investigations about the confusions/surprises generated by the Dirac's notations were
 started only recently  \cite{19} . Probably that further efforts on the line of such 
 investigations will bring a new light on the conditions \eqref{eq:24} as well as on 
 some other QM questions
 
 \section{Beyond the problem of angular observables}
 
 In the previous sections we did a reevaluation of the RSUR \eqref{eq:1}in its role  
 of normality reference element for angular observables. But, as it is known, the
  respective role is a piece of the TIUR (traditional interpretation of uncertainty
  relations) which is still largely present in the nowadays conceptions about QM.
  Then a re-examination of the TIUR global validity  becomes of a direct interest.
  
  The alluded re-examination requires firstly a brief presentation of the TIUR doctrine. 
  In the main the respective doctrine is connected with the preoccupation for giving an 
  unique  and generic interpretation for the RSUR
  \begin{equation}\label{eq:32}
  \Delta A\cdot \Delta B \ge \frac{1}{2}\;| {\langle {[ 
{\hat A,\hat B} ]} \rangle } |
\end{equation}
and for the thought-experimental (\emph{te}) relations
  \begin{equation}\label{eq:33}
\Delta _{te} A \cdot \Delta _{te} B\ge \hbar 
\end{equation}
The relations \eqref{eq:32} were introduced through the mathematical formalism of QM. On
the other hand the relations \eqref{eq:33}, regarding  the \emph{ te}- uncertainties
$\Delta _{te} A $ and $\Delta _{te}B$, were proposed by means of some so called thought 
(or mental) experiments. A bibliography of the significant publications on the history 
of debates about the relations \eqref{eq:32} - \eqref{eq:33}can be found in the works 
\cite{7,20,21,22,23,24,25}. Related to the interpretation of the alluded 
relations it was 
promoted a whole doctrine known as TIUR (traditional interpretation of uncertainty 
relations). Due to the respective doctrine the relations \eqref{eq:32}-\eqref{eq:33}
have a large popularity, they being frequently regarded as crucial physical formulas
or \cite{25} even as expression of \emph{"the most important principle of the
 twentieth century physics"}. But, as a strange aspect, in its partisan literature TIUR is 
often presented so fragmentary and esoteric that it seems to be rather a dim conception but not a 
well-delimited doctrine. However, in spite of such an aspect, from the mentioned 
literature one can infer that in fact TIUR is reducible to the following set of main 
assertions (\textbf{Ass.}) :
\begin{itemize}
\item \textbf{Ass.1}: The quantities $\Delta A$ and $\Delta_{te} A$ from the relations 
\eqref{eq:32} and \eqref{eq:33}have similar significance of measurement uncertainty 
for the quantum observable $A$. Consequently the respective 
relations have also the same generic significance and regard the simultaneous 
measurements of observables $A$ and $B$ 
\item \textbf{Ass.2}: For an observable $A$ considered alone the the quantity $\Delta A$ can
be indefinitely small (even null) 
\item \textbf{Ass.3}: For two observables $A$ and $B$ , considered in simultaneous measurements,
the quantities $\Delta A$ and $\Delta B$  are interrelated through RSUR \eqref{eq:32}
considered as reference formula. So in the non-commutative cases 
($[ {\hat A\,,\,\hat B} ]\; \ne \;0$) the respective quantities cannot be reduced 
concomitantly because their product $\Delta A \cdot \Delta B$  is lower limited by a 
non-null term which depends prevalently on $\hbar$ . On the other hand in the commutative 
cases ($[ {\hat A\,,\,\hat B} ]=0$)  $ \Delta A$ and $\Delta B$ are 
mutually independent, they being allowed to take simultaneously indefinitely small (even
null) values.
\item \textbf{Ass.4 }: The relations \eqref{eq:32} and \eqref{eq:33} are typically QM formulas
and they, as well as the Planck's constant $\hbar$  , have not analogues in classical 
(non-quantum) physics.
\end{itemize}

Now it is clearly that the announced re-examination of the global validity of TIUR  can
 be materialized by scrutinizing the correctness of the assertions \textbf{Ass.1-4}.
 on the line of such scrutiny we note the following remarks (\textbf{Rem.}):
 \begin{itemize}
\item \textbf{Rem.1}: First of all we note that the \emph{te}-relations \eqref{eq:33} are improper
as a reference element for a supposed solid doctrine like TIUR. This because the respective 
relations have only a transitory character due to the fact   that they were founded on
old resolution criteria (introduced by Abbe and Rayleigh - see \cite{26}). But in in 
modern experimental physics  \cite{27,28,29,30,31,32,33,34} some super-resolution 
techniques that overstep the respective criteria are known. Then it is possible to imagine  some
super-resolution-thought-experiments (\emph{srte} ) which instead of \eqref{eq:33} can promote
 the \emph{srte}-relations like:
\begin{equation}\label{eq:34}
  \Delta_{srte}A \cdot \Delta_{srte}B < \hbar
  \end{equation}
  for the \emph{srte}-uncertainties  $ \Delta_{srte}A $ and $\Delta_{srte}B$.The alluded 
  possibility invalidate the assertion \textbf{Ass.1} and incriminates TIUR in connection 
  with one of its main points. 
\item  \textbf{Rem.2}: From \textbf{Rem.1} it results directly that for the debates about 
 TIUR only  the RSUR \eqref{eq:32} remains of interest. But, as it was pointed out in the 
 previous sections, the RSUR \eqref{eq:32} is only a secondary relation, derivable in 
 well-precised conditions from the primary Schwarz formula \eqref{eq:23}. Then it results 
 that in fact TIUR is confronted with a formula which is not consonant with its 
 assertion \textbf{Ass.3}. 
\item \textbf{Rem.3}: Now let us refer to the term "\emph{uncertainty}" used by TIUR for
 quantities like $\Delta A$ from \eqref{eq:32}. We think that the respective term is 
 groundless because of the following facts. As it is defined in the mathematical
  framework of QM the quantity $\Delta A$ signifies a probabilistic estimator 
(standard deviation) of the observable $A$ regarded as a random variable. The mentioned 
framework deals with theoretical concepts and models about the intrinsic (inner) 
properties of the considered system but not with the elements referring to the (possible)
measurements performed on the respective system.  Consequently, for a physical system, 
$\Delta A$ refers to the intrinsic characteristics, reflected in the fluctuations 
(deviations from the mean value) of the observable $A$. Moreover, as the 
expressions \eqref{eq:6} and \eqref{eq:11} suggest, for a system in a 
given state $\Delta A$ has a well 
defined value, connected with the corresponding wave function. The respective value 
cannot be related  with the modifiable evaluations (e.g. by independent or interdependent
reductions) assumed by \textbf{Ass. 2-3}.
\item \textbf{Rem.4}: The alluded modifiable evaluations can be associated with the measurements 
errors/uncertainties, due to the possible changes of the accuracy for the measuring 
devices and procedures. But, as a general rule, such changes regard all the characteristics 
of a random observable $A$ - i.e the mean value $\langle A \rangle $ and 
fluctuation estimators (like $\Delta A $). Moreover such evaluations refer  to all the  
random observables of both quantum and classical type, without differences of principle.
Also, according to the real practice of experimental physics, one can state that for 
avoiding the damages (misconceptions) the descriptions of measurements must not pertain 
to QM or to other chapters of actual theoretical physics. Such a statement is consonant with 
the thinking that \cite{35} : \emph{"in fact the word ('\emph{measurement}') has 
had such a damaging 
effect on the discussions, that ... it should be banned altogether in quantum mechanics"}.
In the spirit of the mentioned statement and thinking the QM, as well as the whole 
theoretical physics, must be concerned only with the (conceptual and mathematical)
models of the intrinsic properties for physical systems. But such a concern disagrees 
with the TIUR's assertions \textbf{Ass. 1-4}.
\item \textbf{Rem.5}: As we have shown in sections 2 and 3 the angular observables 
$L_z - \varphi$ and $N - \phi$ imply situations which are in discordance with 
\textbf{Ass.3}. Surprisingly, similar situations are encountered even for commutable
 observables . Such is the case of Cartesian coordinates $x$ and $y$ regarding a 
 microparticle in a bi-dimensional potential well with inclined walls in respect with 
 the $x - y$ axes. In the respective case \cite{14} the product 
 $\Delta x \cdot \Delta y$ is a non-null quantity, with precisely defined values for
 $\Delta x$  and $\Delta y$. So one finds another example which disaccords 
 with \textbf{Ass.3}.
\item  \textbf{Rem.6}: As it is known TIUR promoted the idea that two observables $A$ and $B$
 to be denoted with the terms "compatible" respectively "incompatible" subsequently of 
 the fact that their operators are commutable ($[ {\hat A,\hat B}]=0$) 
 or not ($[ {\hat A,\hat B} ] \ne 0$). The
  mentioned terms are directly connected with the suppositions of TIUR about the lower 
  limit of the product $\Delta A \cdot \Delta B$ . But it is easy to see that the facts
   presented in the \textbf{Rem.5} prove the desuetude of the mentioned idea. Particularly
   the respective idea becomes self-contradictory for the pairs of observables 
   $L_z$ and$\varphi$
   respectively $x$ and $y$ which ought to be both "compatible" 
   and "incompatible".
\item \textbf{Rem.7}: The quantities $\Delta A$ and $\Delta B$ from RSUR \eqref{eq:32}/
  \eqref{eq:1} are second order probabilistic estimators, evaluated for the same moment 
  of time. Consequently RSUR is a simple uni-temporal probabilistic formula. But the
  respective formula is generalizable in form of some extended relations referring also 
  to the second order probabilistic estimators. So one obtains \cite{14,16}
 bi-temporal, many-observables respectively  quantum-macroscopic relations. For the 
 mentioned extended relations TIUR has to give an interpretation concordant with its own
  essence, if it is a well-grounded doctrine. But to find such an interpretation on 
  natural ways (i.e. without esoteric and/or non-physical considerations) seems to be 
  a difficult (even impossible ) task. In this sense it is significant to remind the lack 
  of success  connected with the above alluded quantum-macroscopic relations. In order 
  to adjust the respective relations to the TIUR's assertions it was resorted to the 
  so called "macroscopic-operators"(see \cite{36}and references). But in fact \cite{14,16}
  the mentioned resort does not ensure for TIUR the avoidance of the involved shortcomings.
  Moreover the respective "macroscopic-operators" are only fictitious concepts     without
   any real applicability in physics. It is also interesting to observe that, in the 
   last decades, the problem of the "macroscopic-operators" and related relations is 
   eschewed in the literature regarding the UR.
\item \textbf{Rem.8}: In classical physics for observables with random character an non-trivial
interest can present also higher order estimators (correlations) \cite{37,38}. This fact 
suggests that in the case of quantum observables, additionally to the second order 
estimators (like $\Delta A=\left( {\delta \hat A\psi ,\delta \hat A\psi } 
\right)^{1/2} $ and 
$(\delta \hat A \psi ,\delta \hat B\psi )$  from \eqref{eq:23}) can be used also  
the higher order correlations such as $((\delta \hat A)^r \psi ,
(\delta \hat B)^s \psi )$) with $r + s \geq 3$. Then, naturally for the respective 
correlations TIUR has to give an interpretation incorporable in his own doctrine.
But it seems to be less probable (or even excluded) that such an interpretation can be 
promoted through  credible arguments.
\item \textbf{Rem.9}: In contradiction with \textbf{Ass.4}  in non-quantum physics 
\cite{14,39,40} there are really some classical formulas that are completely similar 
with the quantum relations \eqref{eq:32} and \eqref{eq:23}. The alluded formulas 
have the form
  \begin{equation}\label{eq:35}
\Delta _{cf} A \cdot \,\Delta _{cf} B \ge \left| {\left\langle {\delta A \cdot 
\delta B} \right\rangle _{cf} } \right|
\end{equation}
Here the standard deviations $\Delta _{cf}A$ and $\Delta _{cf}B$ respectively the
correlation $ {\left\langle {\delta A \cdot \delta B} \right\rangle _{cf}}$ 
refer to the classical fluctuations ($cf$) of the macroscopic observables $A$ and $B$
considered as random variables. Note that in classical conception the fluctuations and 
consequently the relations \eqref{eq:35} regard  the intrinsic (own) properties 
of the macroscopic systems  but not the aspects of the measurements performed on
the respective systems. Then the relations \eqref{eq:35} and \eqref{eq:32} (or \eqref{eq:23})
reveal a classical-quantum similarity. This fact suggests that the quantum quantities
$\Delta A$ and $\Delta B$ from \eqref{eq:32} or \eqref{eq:23} describe intrinsic properties 
(fluctuations) of the quantum observables $A$ and $B$ but not the uncertainties regarding 
the respective observables.
\item \textbf{Rem.10}: The size of the quantities $\Delta A$ and $\Delta_{cf}B$  from
 \eqref{eq:32} and \eqref{eq:35} disclose the level of stochasticity (randomness) for the 
referred observables and systems .On the other 
hand the concrete expressions of $\Delta A$ and $\Delta_{cf}B$ appear \cite{41} as products
between $\hbar$ respectively $k$ (Boltzmann's constant) and quantities which do not contain
$\hbar$ respectively $k$  . So $\hbar$ and $k$ have similar roles of generic indicators 
for stochasticity. But such a similarity disagrees with the TIUR's assertion \textbf{Ass.4}.
\item \textbf{Rem.11}: The TIUR's assertion Ass.3 roused many debates regarding the pair $t -E$
(time - energy) (see \cite{7,42} and references). The alluded debates tried to 
subordinate the description of the pair $t - E$ to the idea that within QM the RSUR 
\eqref{eq:32} is an capital reference element. But as we have shown above the respective 
idea is unjustifiable. Then the direct conclusion is that the mentioned subordination 
is a groundless and unnatural requirement.Such a conclusion can be completed 
with some considerations related both with the above presented discussions and with the 
recent notification \cite{42} that in QM time has a threefold role. In the spirit of the 
mentioned notification the time can be regarded respectively as an external, intrinsic 
or observational entity. The external time $t_{ext}$ \cite{42} \emph{"is identified as the 
parameter entering in the Schr\"{o}dinger equation and measured by an external, detached 
laboratory clock" }. The intrinsic time $t_{int}$ refers to the own properties 
of the quantum objects themselves (such are the spreading of a wave packet, the decay of an unstable state
or the temporal evolution of a quantum observable $A$). In a certain contrast 
with \cite{42} we think that the observational time $t_{obs}$ must be principally associated with the
performances of the measuring devices and procedures (e.g. with the resolution time 
of a device or with the duration of a measuring procedure). With such a view about 
time the announced considerations can be formulated as follows.
\begin{itemize}
\item \textbf{(i)} The external time $t_{ext}$ is a deterministic (non-random or dispersion-free) variable
without fluctuations. Consequently $t_{ext}$ should  not be endowed with an operator nor 
associated with RSUR \eqref{eq:32}. In fact, even such an endowment  leave RSUR \eqref{eq:32}
inapplicable for $t_{ext}$. Indeed , if the operators $t_{ext} = t_{ext}\cdot$ and
$\hat E=i\hbar \frac{\partial }{{\partial t_{ext} }}$ (often promoted in 
publications) are used, one obtains:
 \begin{equation}\label{eq:36}
(\hat E\psi ,\hat t_{ext} \psi )= (\psi ,\hat E \hat t_{ext} \psi )
 - i\hbar 
\end {equation}
With this relation one finds a violation of the conditions \eqref{eq:24} and, consequently, 
a proof of inadequacy for RSUR \eqref{eq:32}/\eqref{eq:25}in respect with the operators 
$\hat t_{ext}$ and $\hat E$ . However , independently of \eqref{eq:36}, for the respective 
operators a relation of \eqref{eq:23}-type   is true i.e. :
\begin{equation}\label{eq:37}
\Delta E \cdot \Delta t_{ext}  \ge \left| {\left( {\delta \hat E \psi,
\delta \hat t_{ext} \psi } \right)} \right|
\end{equation}
But this relation reduces itself to the trivial equality $0 = 0$ (because 
$\left\langle {t_{ext} } \right\rangle =t_{ext} , \delta \hat t=0$
and $\Delta t_{ext} = 0$). Such an equality signifies that, in fact , in the QM framework 
$t _{ext}$ is a deterministic variable in the mentioned sense. In the same framework the 
energy is a random quantity described by the Hamiltonian operator $\hat H$ ( which can be
substituted by $\hat E=i\hbar \,\frac{\partial }{{\partial t_{ext} }}$ due to 
the Schrodinger equation).
\item \textbf{(ii)}For the role of intrinsic time $t _ {int}$ a lot of diverse quantities  were 
promoted 
(see \cite{42}and references). They describe various temporal characteristics of quantum 
systems and each of them is associated with a corresponding time-energy UR. In the main 
the respective relations are manipulated so as to take the generic form : 
\begin{equation}\label{eq:38}
\tau  \cdot \epsilon _{w} \ge \frac{\hbar }{2}
\end{equation}
where $\tau$ denote a characteristic time interval while $\epsilon _{w}$ signify
an energetic 
width. By means of the relation \eqref{eq:38} one expects to harmonize the pair time-energy 
with TIUR and especially with RSUR \eqref{eq:32} regarded as a capital physical formula.
But, on the one hand, from a mathematical and/or physical perspective, the relations
\eqref{eq:38} are not assimilable with RSUR \eqref{eq:32}. On the other hand, as it was proved 
above, in fact TIUR is an unjustified doctrine and RSUR \eqref{eq:32} is not at all a 
capital formula. Then it results that the expectations connected with the relations
\eqref{eq:38} have not a viable object and the respective relations appear as formulas
without a major significance \\
\item \textbf{(iii)} The observational time $t_{obs}$ is dependent outstandingly on the modifications 
(choices or changes ) of the measurement characteristics for the same measured system
in a given state. On the other hand for the respective  system QM associates entities
 (wave function, mean values and standard deviations of observables) which are 
 independent on the mentioned modifications. Such connections with  the measurements 
 characteristics  suggest that the observational times must be described not in the 
 framework of QM but inside of a scientific approach which is distinct and additional 
 in respect with QM. In our opinion \cite{43}an approach of the mentioned kind is required 
 also by the general description of the measurements for for quantum systems.
 \end{itemize}
 \end{itemize}
 
 \section {Conclusions}
 
 The above presented discussions can be summarized through the following concluding
 remarks (\textbf{C.Rem.}):
 \begin{itemize}
\item   \textbf{C.Rem.1}: The RSUR \eqref{eq:1}/\eqref{eq:26}/\eqref{eq:32} is an 
 incorrect reference element   in respect with  the pairs of quantum
 angular observables  $L_z - \varphi$ and $N - \phi$.
 \item \textbf{ C.Rem.2}: In fact the  RSUR  proves itself to be only a a secondary 
  formula valid in well delimited conditions.
\item \textbf{C.Rem.3} : If instead of  RSUR 
 one appeals to the primary Schwarz formula \eqref{eq:23} the description  of the pairs
 $L_z - \varphi$ and $N - \phi$ can be integrated without problems in the framework
  of usual QM. So in the QM framework the true reference element of normality 
  is the respective 
  formula.
  \item \textbf{C.Rem.4}: Because the RSUR plays also the role of reference element for
  TIUR doctrine, the mentioned reevaluation of RSUR requires directly a minute 
  reconsideration of the respective doctrine. Of course that such a reconsideration
  enlarges the observation that \cite{44} UR \emph{"are probably the most controverted 
  formulae 
  in the whole of the theoretical physics" }.
\item   \textbf{C.Rem.5}: Through the alluded reconsideration one finds that in 
  fact TIUR is nothing but an unjustified doctrine. Such a finding consolidate the statement 
  \cite{45}: \emph{"the idea that there are defects in the foundations of orthodox quantum
   theory is unquestionable present in the conscience of many physicists" }.
 \item \textbf{C.Rem.6}: According to the above findings the RSUR loses its aureola
   of crucial formula. In fact RSUR as well as its primary source \eqref{eq:23}(Schwarz 
   formula) must be legitimated as simple QM relations. They refer to the quantum 
   fluctuations (regarded as intrinsic properties of quantum systems) and have natural 
   analogues in classical physics. The mentioned legitimation confirms, from a more 
   general perspective, the observation \cite{17} that there is nothing especially
   quantum mechanical about UR \emph{per se}.
\item \textbf{C.Rem.7}: Because of the assertions \textbf{Ass.1 -3} the above 
   argued reevaluation of TIUR doctrine requires new and natural approaches regarding 
   the description of quantum measurements.  We opine \cite{43}  that such approaches 
   must be done within a theoretical frame which is distinct and additional  in respect
   with the usual QM
\end{itemize}
   
\newpage
\section*{Acknowledgments}\ 
\begin{itemize}
\item I Wish to express my deep gratitude to those authors, publishing companies 
 and libraries which, during the years, helped me with  copies of some publications
 connected with the problems approached here.
\item The investigations in the field of the present paper benefited partially 
of facilities from the grants supported by the Roumanian Ministry of Education and
 Research.
 \end{itemize}
 \section*{List of abbreviations}

\textbf{Ass.} = assertion\\
\textit{cf} = classical fluctuation\\
CR = circular rotations\\
\textbf{C.Rem} = concluding remark\\
\textit{ext} = external\\
\textit{int} = intrinsic\\
NCR = non-circular rotations\\
\textit{obs} = observational\\
QM = quantum mechanics\\
QTP = quantum torsion pendulum\\
\textbf{Rem.} = remark\\
RSUR = Robertson Schr\"{o}dinger uncertainty relation\\
\textbf{Shc.} = shortcoming\\
\textit{srte}= super-resolution-thought-experimental\\
\textit{te} = thought-experimental\\
TIUR = traditional interpretation of uncertainty relations\\
UR = uncertainty relation(s)\\


\newpage

\begin{thebibliography}{99}
\bibitem{1}D.Judge, Phys.Lett. \textbf{5} (1963) 189.
\bibitem{2}D.Judge, J.T.Lewis, Phys.Lett. \textbf{5} (1963) 190.
\bibitem{3}W.H.Luissel, Phys.Lett.\ textbf{7} (1963) 60.
\bibitem{4}P.Caruthers, H.M.Nietto, Rev.Mod.Phys. \textbf{40} (1968) 411.
\bibitem{5}V.V.Dodonov, V.I.Man'ko,Proc.Lebedeev Phys.Institute \textbf{183} (1987) 5
[English version in "Invariants and Evolution of Nonstationary Quantum Systems"
Ed. M.A.Markov ,Nova Science,NY 1989,3-101]
\bibitem{6}R.Lynch, Phys.Reports \textbf{256} (1995) 367.
\bibitem{7}G.Auletta, Foundations and Interpretation of Quantum Mechanics
,World Scientific, Singapore 2000.
\bibitem{8}J.-P. Pellonpaa, Annales Universitatis Turkuensis (Turku, Finland))
Ser.A, Tom \textbf{288} (2002) 5.
\bibitem{9}Yu.P.Vorontsov.Usp.Fiz.Nauk \textbf{172} (2002) 207. [English version 
Physics-Uspeckhi \textbf{45} (2002) 847]
\bibitem{10}K.Kowalsky, J.Rebielinski, J.Phys.A:Math.Gen. \textbf{35} (2002) 1405.
\bibitem{11}K.Kowalsky, J.Rebielinski, Phys.Lett.A \textbf{293} (2002) 109.
\bibitem{12}S.Dumitru, Epistemological Letters \textbf{15} (1977) 1.
\bibitem{13}S.Dumitru, in Recent Advances in Statistical Physics, Ed.B.Datta 
and M.Dutta, World Scientific, Singapore 1987.
\bibitem{14}S. Dumitru, Rev. Roum. Phys. \textbf{33 }(1988) 11.
\bibitem{15}S. Dumitru, in Quantum Field Theory, Quantum Mechanics and Quantum Optics 
Part 1, Symmetries and Algebraic Structures Ed. V.V. Dodonov and V.I. Man'ko 
,Nova Science, New York 1991.
\bibitem{16}S. Dumitru, arXiv:quant-ph/000413.
\bibitem{17}P.R.Holland The Quantum Theory of Motion ,Cambridge University Press,1993.
\bibitem{18}A.Scardiccho, Phys. Lett. A \textbf{300} (2002) 7.
\bibitem{19} F. Gires, Rep.Prog.Phys \textbf{63} (2000) 1983; arXiv:quant-ph/9907069v2 
21 Dec.2001. 
\bibitem{20} M. Jammer, The Conceptual Development of Quantum Mechanics ,Mc Graw Hill, 
New York 1966.
\bibitem{21} B.S. De Witt, N. Graham Am.J.Phys.\textbf{39} (1971) 724.
\bibitem{22} M.Jammer,  The Philosophy of Quantum Mechanics ,Wiley, New York 1974.
\bibitem{23} D.R.Nilson, Bibliography on the History and Philosophy of Quantum Mechanics 
in " Logic and Probability in Quantum Mechanics" Ed. P.Suppes ,D.Reidel, Dordrecht 1976. 
\bibitem{24} L.E.Ballentine Am.J.Phys. \textbf{55} (1987) 785.
\bibitem{25} H.Martens, Uncertainty Principle - Ph.D. Thesis ,Technical University, 
Eidhoven 1991.
\bibitem{26}M. Born, E.Wolf Principles of Optics, Pergamon Press, Oxford 1968.
\bibitem{27}C.Ryachoudhuri,Found. Phys.\textbf{8} (1978) 845.
\bibitem{28}J. Scheer, T.Gotsch, G.Luning, M.Schmidt, H.Ziegel, 
Found.Phys.Lett.\textbf{2 }(1989) 71.
\bibitem{29} R.J.Croca, Hadronic Journal \textbf{22} (1999) 29.
\bibitem{30} R.J.Croca, Towards a Nonlinear Quantum Physics, World Scientific, 
Singapore 2003 (this work, as well as \cite {29}, discuss the potential implications 
of the performances attained in optical experiments such are those reported in \cite{31,32})
\bibitem{31}D.W.Pohl, W.Denk, M.Lanz Appl.Phys.Lett. \textbf{44} (1994) 651.
\bibitem{32}H.Heiselman, D.W.Pohl Appl.Phys \textbf{58} (1994) 89.
\bibitem{33}M.Dyba, S.W.Hell, Phys.Rev.Lett.\textbf{88} (2002) 163901.
\bibitem{34}D.Mugnai, A.Ranfagni, R.Ruggeri, Phys. Lett. A \textbf{311} (2003) 77.
\bibitem{35}J.S.Bell, Physics World \textbf{3} (1990) 33. 
\bibitem{36}R.Jancel, Foundations of Classical and Quantum Statistical Mechanics,
Pergamon, New York 1973. 
\bibitem{37}S.Dumitru, Phys.Lett. \textbf{35 A} (1971) 78; \textbf{41A} (1972) 321.
\bibitem{38}A.Boer, S.Dumitru, Phys.Rev. E \textbf{66}, (2002) 046116. 
\bibitem{39} S.Dumitru,Physica Scripta \textbf{10} (1974) 101.
\bibitem{40}S.Dumitru, A.Boer, Phys. Rev. E \textbf{64} (2001) 021108. 
\bibitem{41}S.Dumitru, Physics Essays \textbf{6} (1993) 5. 
\bibitem{42}P.Busch, Lecture Notes in Physics \textbf{72M} (2002) 69 (Springer); arXiv: 
quant-ph/0105049v1 
\bibitem{43}S. Dumitru, arXiv: quant-ph/0307008 (next preprint).
\bibitem{44}M.Bunge, in: Denken und Umdenken (zu Werk und Werkung von Werner Heisenberg),
Ed. H.Pfeifer, Piper, Munchen 1977. 
\bibitem{45}C.Piron, Lecture  Notes in Physics \textbf{153} (1982) 179 (Springer). 
 \end{thebibliography}
\end{document}